\documentclass[preprintnumbers,twocolumn,showpacs,amsmath,amssymb]{revtex4}

\usepackage{epsfig}
\usepackage{graphicx}
\usepackage{dcolumn}
\usepackage{bm}

\newcommand{\be}{\begin{equation}}
\newcommand{\ee}{\end{equation}}

\newcommand{\bea}{\begin{eqnarray}}
\newcommand{\eea}{\end{eqnarray}}
\newcommand{\bfig}{\begin{figure}}
\newcommand{\efig}{\end{figure}}
\newcommand{\bc}{\begin{center}}
\newcommand{\ec}{\end{center}}
\newcommand{\TeV}{\unskip\,\mathrm{TeV}}
\newcommand{\GeV}{\unskip\,\mathrm{GeV}}
\newcommand{\MeV}{\unskip\,\mathrm{MeV}}

\newcommand{\Pe}{\mathrm{e}}

\newcommand{\PZ}{\mathrm{Z}}
\newcommand{\PW}{\mathrm{W}}

\def\mathswitch#1{\relax\ifmmode#1\else$#1$\fi}
\def\mathswitchr#1{\relax\ifmmode{\mathrm{#1}}\else$\mathrm{#1}$\fi}

\newcommand{\MW}{\mathswitch {M_\PW}}
\newcommand{\PH}{\mathswitchr H}

\newcommand{\MZ}{\mathswitch {M_\PZ}}
\newcommand{\MH}{\mathswitch {M_\PH}}
\newcommand{\Me}{\mathswitch {m_\Pe}}

\newcommand{\Mt}{\mathswitch {m_\Pt}}
\newcommand{\GW}{\Gamma_{\PW}}

\newcommand{\GZ}{\Gamma_{\PZ}}

\newcommand{\LEP}{{\mathrm{LEP}}}

\newcommand{\Pt}{\mathswitchr t}

\def\Ga{\Gamma}
\def\d{\hbox{d}}
\def\OO{y}
\begin{document}
\preprint{PSI-PR-09-09, ZU-TH 08/09}

\title{Electroweak corrections to three-jet production 
in \boldmath{$\Pe^+\Pe^-$} annihilation}

\author{Ansgar Denner$^a$, Stefan Dittmaier$^{b,c}$, Thomas Gehrmann$^d$,
Christian Kurz$^{a,d}$}
 \affiliation{$^a$ Paul Scherrer Institut, CH-5232 Villigen PSI,
Switzerland\\
$^b$ Physikalisches Institut, Albert-Ludwigs-Universit\"at
Freiburg, D-79104 Freiburg, Germany\\
$^c$ Max-Planck-Institut f\"ur Physik
(Werner-Heisenberg-Institut), D-80805 M\"unchen, Germany\\
$^d$ Institut f\"ur Theoretische Physik,
Universit\"at Z\"urich, CH-8057 Z\"urich, Switzerland
}

\date{\today}

\begin{abstract}
We compute the electroweak ${\cal O}(\alpha^3\alpha_{\mathrm{s}})$ corrections to
three-jet production and related event-shape observables at electron--positron 
colliders. We properly account for the experimental photon isolation criteria
and for the corrections to the total hadronic cross section. Corrections to 
the three-jet rate and to normalised event-shape distributions turn out 
to be at the few-per-cent level. 
\end{abstract}

\maketitle


Precision QCD studies at electron--positron colliders rely on the
measurement of the three-jet production cross section and related
event-shape observables.  The deviation from simple two-jet
configurations is proportional to the strong coupling constant
$\alpha_{\mathrm{s}}$, so that by comparing the measured three-jet
rate and related event shapes (see, e.g., Ref.~\cite{dissertori}) 
with the theoretical
predictions, one can determine $\alpha_{\mathrm{s}}$. Including
electroweak coupling factors, the leading-order (LO) contribution to
this process is of order $\alpha^2\alpha_\mathrm{s}$.

Owing to recent calculational progress, the QCD predictions for event
shapes~\cite{ourevent,weinzierlevent} and three-jet
production~\cite{our3j,weinzierl3j} are accurate to
next-to-next-to-leading order (NNLO, $\alpha^2\alpha_\mathrm{s}^3$) in
QCD perturbation theory.  Depending on the observable under
consideration, the numerical magnitude of the NNLO corrections varies
between three and twenty per cent. Inclusion of these corrections
results in an estimated residual uncertainty of the QCD prediction
from missing higher orders at the level of below five per cent for the
event-shape distributions, and below one per cent for the three-jet
cross section.  At this level of theoretical precision,
 higher-order electroweak effects could be of comparable
magnitude. At present, only partial calculations of electroweak
corrections to three-jet production and event shapes are
available~\cite{moretti}, which can not be compared with experimental
data directly. In this work, we present the first calculation of the
NLO electroweak ($\alpha^3\alpha_s$) corrections to three-jet
observables in $\Pe^+\Pe^-$ collisions including the
quark--antiquark--photon ($q\bar{q}\gamma$) final states.  Note that
the QCD corrections to these final states are of the same perturbative
order as the genuine electroweak corrections to
quark--antiquark--gluon ($q\bar{q}\mathrm{g}$) final states.  Since
photons produced in association with hadrons can never be fully
isolated, both types of corrections have to be taken into account.

Event-shape measurements at LEP usually rely on a standard set of six
variables $y$, defined for example in Ref.~\cite{alephqcd}: thrust
$T$, $C$-parameter, heavy jet mass $\rho$, wide and total jet
broadenings $B_{\mathrm{W}}$ and $B_{\mathrm{T}}$, and
two-to-three-jet transition parameter in the Durham algorithm $Y_3$.
The experimentally measured event-shape distribution
$$\frac{1}{\sigma_{{\rm had}}}\, \frac{\d\sigma}{\d y}$$ is normalised to the
total hadronic cross section. In the perturbative expansion, 
it turns out to be most appropriate to  consider
the expansion of this ratio, 
which reads to NNLO in QCD and NLO in the electroweak theory
\begin{eqnarray}
\frac{1}{\sigma_{{\rm had}}}\, \frac{\d\sigma}{\d \OO} &=&
\left(\frac{\alpha_\mathrm{s}}{2\pi}\right) \frac{\d \bar A }{\d \OO} +
\left(\frac{\alpha_\mathrm{s}}{2\pi}\right)^2 \frac{\d \bar B }{\d \OO}
+ \left(\frac{\alpha_\mathrm{s}}{2\pi}\right)^3
\frac{\d \bar C }{\d \OO} \nonumber \\ &&
+ 
\left(\frac{\alpha}{2\pi}\right)
\frac{\d \delta_{\gamma}}{\d \OO}
+  \left(\frac{\alpha_s}{2\pi}\right)
\left(\frac{\alpha}{2\pi}\right)
\frac{\d \delta_{\mathrm{EW}}}{\d \OO}\;,
\label{eq:dist}
\end{eqnarray}
where the fact is used that the perturbative expansion of
$\sigma_{{\rm had}}$ starts at order $\alpha^2$.  The calculation of
the QCD coefficients $\bar A$, $\bar B$, and $\bar C$ is described in
Refs.~\cite{ourevent,weinzierlevent}.  
The LO purely 
electromagnetic contribution $\delta_\gamma$ arises from 
tree-level $q\bar q\gamma$ final states without a gluon. 
The NLO electroweak coefficient
$\delta_{\mathrm{EW}}$ receives contributions from the ${\cal
  O}(\alpha)$ correction to the hadronic cross section,
\begin{equation}
\sigma_{{\rm had}} = \sigma_0 \left[ 1 + 
\left(\frac{\alpha}{2\pi}\right) \delta_{\sigma,1} \right],
\label{eq:sighad}
\end{equation}
and from the genuine ${\cal O}(\alpha^3\alpha_s)$ contribution to the 
event-shape distribution
$$
\frac{1}{\sigma_0}\, \frac{\d\sigma}{\d \OO} =
\left(\frac{\alpha_\mathrm{s}}{2\pi}\right) \frac{\d \bar A }{\d \OO} +
\left(\frac{\alpha_\mathrm{s}}{2\pi}\right)
 \left(\frac{\alpha}{2\pi}\right)
\frac{\d \delta_{\bar A}}{\d \OO}\,,
$$ 
such that 
\begin{equation}
\frac{\d \delta_{\mathrm{EW}}}{\d \OO} = \frac{\d \delta_{\bar A}}{\d \OO} 
- \delta_{\sigma,1} \frac{\d \bar A }{\d \OO}
\end{equation}
yields the full NLO electroweak correction. Both terms are to be
evaluated with the same event-selection cuts.  As shown in the
following, many of the numerically dominant contributions, especially
from initial-state radiation, cancel in this difference.

In the experimental measurement of three-jet observables
at electron--positron centre-of-mass energy $\sqrt{s}$, several cuts are
applied to reduce the contributions from photonic radiation. In our
calculation, we apply the criteria used in the ALEPH analysis~\cite{alephqcd}.
Very similar criteria were also applied by the other experiments~\cite{lepqcd}.
Particles contribute to the final state only if they are 
within the detector acceptance,
defined by the production angle relative to the beam direction,
$|{\cos\theta}|<0.965$. Events are accepted if the reconstructed 
invariant mass squared $s'$ of the final-state particles is larger 
than $s_{{\rm cut}}=0.81 s$.
To reduce the contribution from hard photon radiation, the final-state  
particles  are clustered into jets using the Durham algorithm with resolution 
parameter $y_{{\rm cut}}=0.002$. If one of the resulting jets contains a 
photon carrying a fraction $z_{\gamma}>z_{\gamma,{\rm cut}}=0.9$ 
of the jet energy, 
it is considered to be an isolated photon, and the event is discarded. 
The event-shape variables are then computed in the centre-of-mass frame of 
the final-state momenta, which can be boosted relative to the 
$\Pe^+\Pe^-$ centre-of-mass frame, if particles are outside the detector 
acceptance. 

In the computation of the ${\cal O}(\alpha)$ corrections to the total
hadronic cross section, we include the virtual electroweak corrections
to $q\bar q$ final states, and the real radiation corrections from
$q\bar q \gamma$ final states, provided the above event-selection
criteria are fulfilled.  The corrections to the event-shape
distributions receive contributions from the virtual electroweak
corrections to the $q\bar q \mathrm{g}$ final state, the virtual QCD
corrections to the $q\bar q\gamma$ final state, and from the real
radiation $q\bar q \mathrm{g}\gamma$ final state.  To separate the
divergent real radiation contributions,  we used both the dipole
subtraction method~\cite{cs,dittmair} and phase-space
slicing~\cite{slice}, resulting in two independent implementations. 
Soft singularities are present in the virtual
and real corrections. They are regularized dimensionally or with
infinitesimal photon and gluon masses, and cancel in the sum.
Collinear singularities from photon radiation off the incoming leptons
(initial-state radiation, ISR) are only partially cancelled. The
left-over collinear ISR singularity is regularized by the electron
mass and absorbed into the initial-state radiator function, which we
consider either at fixed order, or in a leading-logarithmic (LL)
resummation~\cite{Beenakker:1996kt}. Owing to the specific nature of
the event selection, also photon radiation off the outgoing quarks
(final-state radiation, FSR) is only partially cancelled.  The
left-over FSR singularity arises from the isolated photon definition,
which vetoes on photon jets with $z_{\gamma}>z_{\gamma,{\rm cut}}$.
This singularity is absorbed into the photon fragmentation function,
which we apply in the fixed-order approach of Ref.~\cite{glovermorgan}. For the
non-perturbative contribution to this function, we use the ${\cal
  O}(\alpha)$ two-parameter fit of ALEPH~\cite{alephfrag}. The fragmentation 
contribution derived in Ref.~\cite{glovermorgan} is based on phase space 
slicing and dimensional regularization. We recomputed this contribution 
using subtraction and mass regularization~\cite{dittmair}. 

The Feynman diagrams for the virtual corrections are generated with
{\sc  FeynArts}~\cite{Kublbeck:1990xc,Hahn:2000kx}.  Using two independent
inhouse {\sc Mathematica} routines, one of which builds upon
{\sc FormCalc}~\cite{Hahn:1998yk}, each diagram is expressed in terms
of standard matrix elements and coefficients of tensor integrals. The
tensor integral coefficients  are numerically reduced to standard
scalar integrals using the methods described in
Refs.~\cite{Denner:2002ii,Denner:2005nn}.  The scalar master integrals
are evaluated using the methods and results of
Refs.~\cite{'tHooft:1978xw,Beenakker:1988jr,Denner:1991qq}, where UV
divergences are regularized dimensionally. 
For IR divergences two alternative regularizations are employed, one that is
fully based on dimensional regularization with massless light fermions, 
gluons, and photons, and another that is based on infinitesimal photon 
and gluon masses and small fermion masses. 
The loop integrals are translated from one scheme to the other as described 
in Ref.~\cite{Dittmaier:2003bc}.
 
The Z-boson resonance is described in the complex-mass scheme
\cite{Denner:1999gp,Denner:2005fg}, and its mass is fixed from the
complex pole. The electromagnetic couplings appearing in LO
are parametrized in the $G_\mu$ scheme, i.e., they are fixed
via 
$$
\alpha=\alpha_{G_\mu}=\sqrt{2}G_\mu M_\PW^2 \left(1-M_\PW^2/M_\PZ^2\right)/\pi.
$$
As the leading electromagnetic corrections are related to the
emission of real photons, we fix the electromagnetic coupling appearing
in the relative corrections by $\alpha=\alpha(0)$,
which is the appropriate choice for the leading photonic corrections.
Accordingly the cross section for $\Pe^+\Pe^-\to q\bar q \mathrm{g}$ is
proportional to $\alpha_{G_\mu}^2\alpha_{\mathrm{s}}$ while the
electroweak corrections to this process are proportional to
$\alpha(0)\alpha_{G_\mu}^2\alpha_{\mathrm{s}}$.

We performed two independent calculations of all ingredients resulting
in two independent {\sc Fortran} codes, one of them being an extension of
{\sc Pole} \cite{Accomando:2005ra}.  

We use the following values of the electroweak and QCD parameters: 
\bea
\begin{array}[b]{r@{\,}lr@{\,}l}
G_{\mu} &= 1.16637\times 10^{-5}\GeV^{-2}\hspace{-0.1em},&\MH&=120\GeV \\
\alpha(0) &= 1/137.03599911, &
\alpha_{\mathrm{s}}(\MZ) &= 0.1176,\\                          
\Me &= 0.51099892 \MeV, &  \Mt &= 171.0\GeV.
\end{array}
\eea
Because we employ a fixed
width in the resonant W- and Z-boson propagators in contrast to the
approach used at LEP to fit the W~and Z~resonances, where running
widths are taken,  we have to convert the ``on-shell''
values of $M_V^{\LEP}$ and $\Ga_V^{\LEP}$ ($V=\PW,\PZ$), resulting
from LEP, to the ``pole values'' denoted by $M_V$ and $\Ga_V$,
leading to \cite{Bardin:1988xt}:
\bea
\begin{array}[b]{r@{\,}l@{\qquad}r@{\,}l}
\MW &= 80.375\ldots\GeV, & \GW &= 2.140\ldots\GeV, \\
\MZ &= 91.1535\ldots\GeV,& \GZ &= 2.4943\ldots\GeV.
\label{eq:m_ga_pole_num}
\end{array}
\eea
In the final state we take all light quarks into account, including b quarks. 
\begin{figure}
\includegraphics{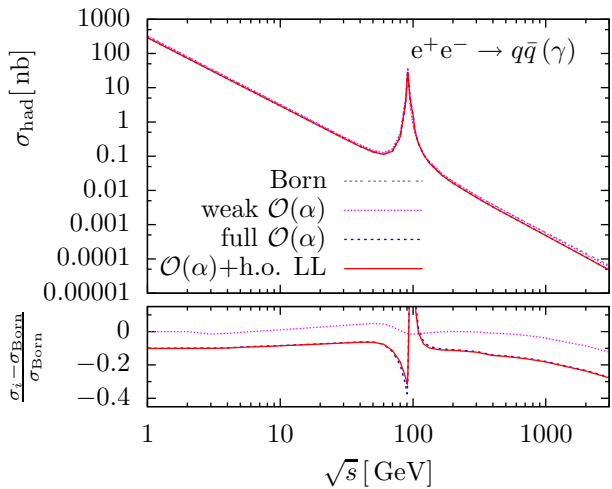}
\caption{Total hadronic cross section $\sigma_{\mathrm{had}}\left(\sqrt{s}\right)$.}
\label{fig:sighad}
\end{figure}

In Fig.~\ref{fig:sighad}, we display the total hadronic cross
section $\sigma_{{\rm had}}$ for the above event-selection criteria
including NLO electroweak corrections and the relative corrections
separately. 
For the latter, ``full'' and ``weak'' refers to the electroweak NLO
corrections with and without purely photonic corrections, respectively,
and ``h.o.~LL'' indicates the inclusion of the higher-order ISR effects.
For most energies, the full ${\cal O}(\alpha)$ corrections are sizable and
negative, ranging between $-30\%$ at the Z peak and about $-10\%$ at
energies above and below. The numerically largest contribution is
always due to ISR. However, above the Z resonance up to about $110\GeV$ the
corrections are positive and of the order of the Born cross section due to the
well-known radiative return phenomenon \cite{Bardin:1989qr} that
occurs in this region because of
our choice of $s_{\mathrm{cut}}$. As it is not relevant
experimentally, we did not fully resolve it in  Fig.~\ref{fig:sighad}.
Below $60\GeV$ and above $120\GeV$ the magnitude of the corrections is increased
due to LL resummation of ISR, whereas it is decreased in the region in
between.
The virtual
one-loop weak corrections (from fermionic and massive bosonic loops)
yield only a moderate correction between $-5$ and $+5$\%. The increase
for energies above $1\TeV$ is due to electroweak Sudakov logarithms.
\begin{figure}
\includegraphics{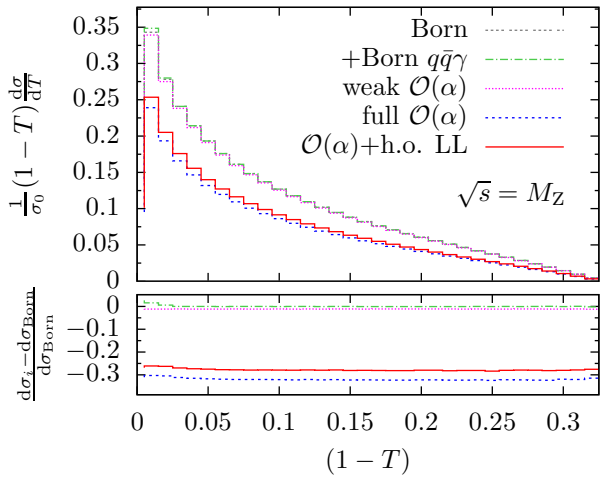}
\caption{Differential thrust distribution
at $\sqrt{s}=M_{\mathrm{Z}}$.}
\label{fig:sigthrust}
\end{figure}

Using the same event-selection cuts, 
Fig.~\ref{fig:sigthrust} displays the differential thrust distribution
at $\sqrt{s}=M_{\mathrm{Z}}$, including NLO electroweak contributions. 
The distributions are weighted by $(1-T)$, evaluated at each bin centre.
The Born
contribution is given by the $\overline{A}$-term of (\ref{eq:dist}), while 
the full ${\cal O}(\alpha)$ corrections contain the tree-level 
$q\bar q \gamma$ contribution $\delta_\gamma$ and the NLO electroweak 
contribution $\delta_{\overline{A}}$. Again, we observe 
large negative corrections due to ISR, and moderate weak
corrections.  The corrections are largely constant for $T<0.95$, where 
the isolated photon veto rejects all contributions from 
$q\bar q\gamma$ final states. For $T>0.95$, corresponding to the two-jet limit,
we find a substantial contribution from 
$q\bar q\gamma$ final states already at LO ($\alpha^3$). Moreover, it
turns out that the electromagnetic corrections depend non-trivially on the 
event-selection cuts. 

\begin{figure}
\includegraphics{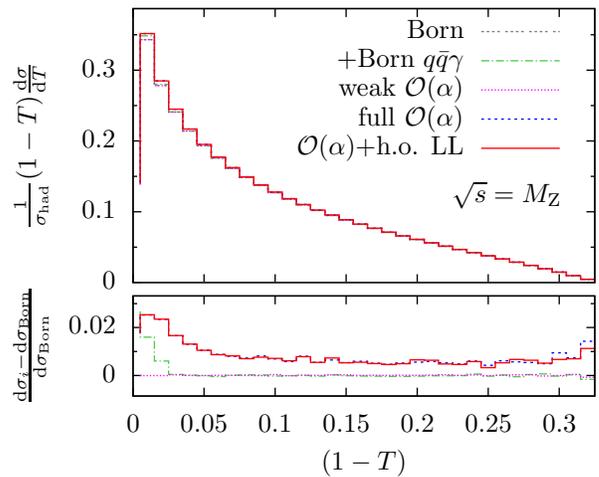}
\caption{Differential thrust distribution
at $\sqrt{s}=M_{\mathrm{Z}}$ normalised to $\sigma_{\mathrm{had}}$.}
\label{fig:sigthrustnorm}
\end{figure}
In expanding the corrections according to (\ref{eq:dist}), and
retaining only terms up to LO in $\alpha_s$, we obtain the
genuine electroweak corrections to normalised event-shape
distributions, which we display for thrust at $\sqrt{s}=M_\mathrm{Z}$ in
Fig.~\ref{fig:sigthrustnorm}. Again, the Born contribution is 
given by the $\overline{A}$-term, while the ${\cal O}(\alpha)$ corrections
now consist of  $\delta_\gamma$ and $\delta_{{\rm EW}}$. It can be seen very 
clearly that the
large ISR corrections cancel between the event-shape distribution and
the normalisation to the hadronic cross section, resulting in electroweak
corrections of a few per cent. Moreover, effects from ISR
resummation are largely reduced as well.
The purely weak corrections are below 0.5 per mille.

\begin{figure}
\includegraphics{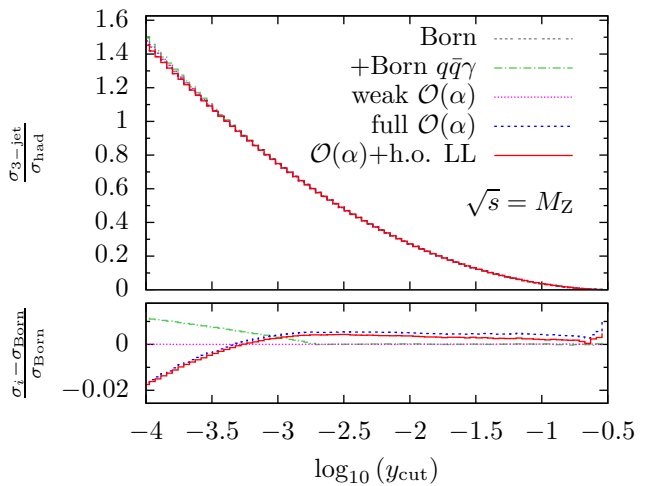}
\caption{Three-jet rate
at $\sqrt{s}=\MZ$ normalised to $\sigma_{\mathrm{had}}$.}
\label{fig:sig3jet_MZ}
\end{figure}
As shown in Fig.~\ref{fig:sig3jet_MZ}, we observe a similar behaviour for the
three-jet rate in the region $y_\mathrm{cut}\gtrsim 0.002$. For $y_\mathrm{cut}\lesssim 0.002$,
$q\bar q\gamma$ final states contribute and lead to a distortion of the shape. For 
$y_\mathrm{cut}\lesssim 0.0005$ the three-jet rate becomes larger than 
$\sigma_{\mathrm{had}}$. In this region, fixed order perturbation theory 
becomes unreliable due to large logarithmic corrections at all orders. 
By improving the LO QCD prediction used here with  
NLO and NNLO QCD corrections, which are 
large and negative 
\cite{our3j,weinzierlevent} for small $y_\mathrm{cut}$,
it is possible to 
extend the range of validity of the fixed order predictions. \\
In Ref.~\cite{moretti}, another calculation of electroweak corrections to 
three-jet observables was performed, which  differs in two important aspects 
from the work presented here. It considered only the corrections to 
$q\bar q \mathrm{g}$ final states, while $q\bar q \gamma$ final states at
 LO and NLO were not taken into account. To remove singularities 
 associated with infrared gluons in $q\bar q \gamma\mathrm{g}$ final states,
  event-shape observables were calculated from the reconstructed jet momenta and 
 not from the parton momenta, as used in experiment and in our work.
Moreover, the NLO electroweak corrections to the hadronic cross section 
were not taken into account, such that only unnormalised 
distributions were considered. Owing to these 
substantial differences, a direct comparison with the results
of Ref.~\cite{moretti} is not possible. Taking care of the different
renormalizaion of $\alpha$, we do observe, however, in the
unnormalized  distribution, Fig.~\ref{fig:sigthrust}, that the relative 
size of the ${\cal O}(\alpha)$ weak and exact corrections, and of the 
LL-improved corrections to the thrust distribution agree at the per-cent 
level with the results of Ref.~\cite{moretti}, except in the region 
$(1-T)<0.05$, where $q\bar q \gamma$ final states contribute.

Data on event-shape distributions and jet cross sections have been corrected 
for photonic radiation effects modelled
by standard LL parton-shower Monte Carlo
programs. They can thus not be compared directly with the NLO electroweak
corrections computed here. Incorporation of these corrections requires a more 
profound reanalysis of LEP data, in order to quantify the impact of the 
NLO electroweak corrections on precision QCD studies, such as the 
precise extraction of the strong coupling constant at
 NNLO in QCD~\cite{asnnlo}.
 
Acknowledgement: This work was supported in part by the Swiss National Science
Foundation (SNF) under contracts 200020-116756 and 200020-117602 and by the
European Community's Marie-Curie Research Training Network HEPTOOLS under contract
MRTN-CT-2006-035505.

\end{document}